\documentstyle{article}
\newcommand{\be}{\begin{equation}}
\newcommand{\ee}{\end{equation}}

\begin{document}
\begin{center}
{\Large\bf The Evolving Universe and the Puzzling Cosmological Parameters}\\
\vspace{0.8in}
{\large Arbab I. Arbab\footnote{arbab64@hotmail.com}}\\
\vspace{0.3in}
{\small Department of Physics, Faculty of Science, University of Khartoum,
Khartoum 11115, SUDAN}
\end{center}
\vspace{1in}
\centerline{ABSTRACT}

The universe is found to have undergone several phases in which the gravitational
constant had different behaviors.
During some epochs the energy density of the universe
remained constant and the universe remained static. In the radiation
dominated epoch the
radiation field satisfies the Stefan's formula while the scale factor
varies linearly with time time. The model enhances the formation of the structure in
the universe as observed today.
\\ \\
\large
{\bf Introduction}
\\  \\
The idea of a variable gravitational constant was first introduced
by Dirac [4] in his Large Number Hypothesis, that the fundamental
constants of physics and astronomy are related by some simple
functions of our present epoch. He thus suggested that the
gravitational constant $G$, as measured in atomic standards,
varies with time as $t^{-1}$ and the number of nucleons in the
universe varies as $t^2$. While Dirac suggestion that $G\propto
t^{-1}$ during the entire evolution of the universe, Brans-Dicke
(BD) [5] theory suggests that the gravitational constant has a
different evolution in different eras. In BD theory the
gravitational constant is related to the inverse of a scalar field
that changes with time. The Brans-Dicke theory represents an
extension of the general theory relativity to include a scalar
particles for the propagation of the gravitational interaction
(long range). However, these particles have not yet been observed.
For a positive coupling between scalars and geometry ($\omega$),
the theory implies a decreasing gravitation constant, though some
application requires this parameter to be negative in th early
universe [8,9]. At the present time observations require $\omega$
to be 500. Models of an increasing gravitational constant with a
presence of a decaying cosmological constant are recently proposed
by Abdel Rahman [2], Beesham [7]. In these models the variation of
the gravitational constant is canceled out by the variation in the
cosmological constant so that  the energy conservation law remains
intact. Though this approach is not covariant yet looks appealing.
Bertolami [8] has found  a similar solution in the BD theory for
the matter dominated present epoch as well as the  radiation
dominated epoch, but with a negative $\omega$. Abdel Rahman has
shown that during the radiation dominated universe the
gravitational constant varies quadratically with time. In the same
line we have generalized this work to include a general variation
of $G$ with time in the presence of a cosmological term having
$\Lambda\propto t^{-2}$ [12]. Thus our model encompasses Dirac's
model as well as Brans-Dicke model. Brans-Dicke and Dirac models
are said to be Machian as they satisfy Mach's principle [11]. It
has been shown that the relation $\frac{GM}{Rc^2}\sim 1$ is valid
for the whole evolution of the universe and represents the Mach's
principle [11]. Employing this relation together with our model we
have been able to discover what the early universe as well as the
present universe looks like. In an  earlier work we have shown
that our G-varying model is consistent with the palaeontological
as well as geophysical data so far known [13]. \\ The unification
recipe was applied to the three interactions excluding gravity. It
depends on the idea that the coupling constants of the three
interaction are running coupling constants (i.e. depend on
energy). The three interactions get unified around the Grand
Unified (GUT) scale where the temperature of the universe was
around $10^{15}\rm GeV$. However, knowledge of the gravitational
interactions at this epoch can only be extrapolated from the
general theory of relativity. The idea of a variable gravitational
constant can be viewed as making the gravitational coupling ($G$)
a running coupling constant (i.e. $G$ is made to  depend on energy
rather than time). Thus one may get some clue about the behavior
of gravity in the very early universe that may not be possible in
the framework of the ordinary general relativity. Though one
believes that the appropriate way for a complete description of
gravity near Planck's scale is the quantum gravity yet this
approach, albeit, phenomenological merits some attention. The
universe is found to evolve into different phases. The model is
found to give a good account for the structure formation in the
universe that we observe today.
\\
\\ {\bf Varying G and the Early Universe}
\\ \\
In an earlier work, we have shown that $G\propto t^2$ and the scale factor
$R\propto t$ in the radiation epoch [1]. Using Mach's principle ($c=1$),
\begin{equation}
\frac{GM}{R}\sim 1
\end{equation}
We obtains
\begin{equation}
\frac{G_P}{G_N}\frac{M_P}{M_N}=\frac{R_P}{R_N}, \ \ \
\frac{G_P}{G_G}\frac{M_P}{M_G}=\frac{R_P}{R_G}, \ \ \
\frac{G_P}{G_W}\frac{M_P}{M_W}=\frac{R_P}{R_W},
\end{equation}
and
\begin{equation}
\frac{R_P}{R_N}=\frac{t_P}{t_N}, \ \ \  \frac{R_P}{R_W}=\frac{t_P}{t_W}, \
\ \
\frac{R_P}{R_G}=\frac{t_P}{t_G}
\end{equation}
and
\begin{equation}
\frac{G_P}{G_N}=(\frac{t_P}{t_N})^2,\ \ \
\frac{G_P}{G_W}=(\frac{t_P}{t_W})^2, \ \
\
\frac{G_P}{G_G}=(\frac{t_P}{t_G})^2
\end{equation}
where, $\rm M_W, M_G, M_N, M_P$ are the masses at Electroweak ($\rm t_W$), GUT
($\rm t_G$), Nuclear ($\rm t_N$), and Planck's time ($\rm t_P$),
and $\rm R_G, R_W, R_P$ are the radius of the universe at GUT, Electroweak and
Planck time, respectively, $\rm G_G, G_W, G_P$ are the
value of the gravitational constant at the GUT, Electroweak and Planck time,
respectively. With $\rm M_P=10^{-5}\rm g, R_P=10^{-33}\rm
cm, t_P=10^{-43}\rm s, t_N=10^{-23}\rm s$, one obtains
\begin{equation}
G_P=G_0, \ G_N=10^{40} G_0, \  G_G=10^6G_0,  \ G_W=10^{32}G_0,
\end{equation}
\begin{equation}
M_N=10^{-1}\rm GeV, \ M_G=10^{16}\rm GeV, \  M_W=10^3\rm GeV
\end{equation}
and
\begin{equation}
R_G=10^{-30}\rm cm, \ \ R_W=10^{-17}\rm cm
\end{equation}

See Table (1) for a complete enumeration. According to the above equations
one obtains
\begin{equation}
t_G=10^{-40} s, \ \ t_W=10^{-27} s
\end{equation}
Though the time for the GUT and Electroweak are different from the
Standard Model the radius of the universe is the same at these epochs.
A time $t=10^{-35}\rm sec$ would correspond to a mass $M=10^{11}\rm GeV$.
Notice that the relation  $G\propto t^2$
is consistent with the above formula.
In fact one observes that the Planck's mass is evolving with time. Thus one
can say that the $M_G,M_W, M_N$ are manifestation of this mass in
different epochs of the evolution of the universe. Only at Planck's time
the Planck's mass has the value $10^{-5}g$. The effect of the Planck's mass
evolution is that gravitational strength changes with time too.
And sine $G\propto t^2$, eq.(1) gives $M\propto t^{-1}$. This result shows that the
matter is annihilated continuously as the universe expands. Thus it is
clear that this
annihilated mass will contribute to the increasing of the entropy.
\\
It is clear that the universe in this model does not have a particle horizon or
monopole problem, that exist in the standard model.
We see that the gravitational constant
had a very different (and big) value in the early universe.
One also anticipates that the universe to enter an epoch characterized by
\begin{equation}
G\propto t^{-2},\ \ R= \ \ \rm const., \ \ \rho= \ \  const
\end{equation}
This solution is obtainable from BD theory with $\omega=-1$. It is generally known
that  this
solution corresponds to a low energy string model.
This period is essential to bring the value of the gravitational constant down to a
reasonable value before nucleosynthesis starts. In this static universe the
perturbations grows exponentially which will later  become the seeds for formation
of structure in the universe. In the matter dominated (MD) phase one has
\begin{equation}
G=G_0(\frac{t}{t_0}), \ \ R=R_0(\frac{t}{t_0})
\end{equation}
where the subscript '0' denotes present day quantity. This variation
law, together with eq.(1), gives, M=const. Thus the matter creation stopped
by the advent of the matter domination. This situation continues up to the
present epoch.
From eq.(1) one can write
\begin{equation}
\frac{G_0}{G_P}\frac{M_0}{M_P}=\frac{R_0}{R_P},\ \
\frac{G_0}{G_N}\frac{M_0}{M_N}=\frac{R_0}{R_N},\ \
\frac{G_0}{G_W}\frac{M_0}{M_W}=\frac{R_0}{R_W},\ \
\frac{G_0}{G_G}\frac{M_0}{M_G}=\frac{R_0}{R_G}
\end{equation}
which give the same results as obtained from eqs.(5)-(8), and predict that
the mass of the present universe is $\sim 10^{56}\rm g$ ! Thus we have shown that
eq.(1) is very fundamental and gives us a lot of clues to our
evolving universe.
\\ \\
{\bf Formation of galaxies}
\\ \\
The development of inhomogeneities in the universe is described by
the density contrast $\delta\equiv\frac{\delta\rho}{\rho}$. Since
$\rho\propto R^{-4}$ in the radiation dominated (RD) phase
($t<t_{eq}$) and $\rho\propto R^{-3}$ in the matter dominated (MD)
phase ($t>t_{eq}$) we get (where $t_{eq}$ is the time when matter
and radiation were in equilibrium) [3] $$
\frac{\delta\rho}{\rho}\propto \left\{ \begin{array}{ll}
  R  \ \ \  & \ t < t_{eq}\\
   R \ \ \ & \ t > t_{eq}\\
\end{array} \right \}
$$
It clear that the density contrast is the same in both radiation dominated and
matter
dominated phase. The  amplitude of mode with $\lambda > d_H$ always grows, where
$\lambda$ is the wavelength of the perturbation and $d_H$ is the Hubble radius. The
future growing modes after entering the Hubble radius($\lambda < d_H $)  depend on
the
pressure distribution of matter which may prevent gravitational enhancement of the
density contrast. This is characterized by the Jeans length $\lambda_J$ where [3,6]
\begin{equation}
\lambda_J=\sqrt{{\pi}}\frac{v}{(G\rho)^{1/2}}
\end{equation}
($v$ is the velocity dispersion of the perturbed component) and $\rho$ is the
density of the component which is most dominant gravitationally, i.e. the one that
makes the perturbation collapse.
For all $\lambda>>\lambda_J$ in the matter dominated phase, the amplitude grows as
$R$. Note that $v\propto R^{-1}$ due to the red shifting of the momentum
$p\propto mv\propto R^{-1}$. One can now define the Jeans's mass as [3,6]
\begin{equation}
M_J=\frac{4\pi}{3}\rho(\frac{\lambda_J}{2})^3
\end{equation}
Since $\rho\propto R^{-4}$ in the radiation dominated phase and $\rho\propto
R^{-3}$ in the matter dominated phase, we see that
$$
M_J\propto
\left\{ \begin{array}{ll}
 R^{-1} \ \ & \ R < R_{eq}\\
 R^{-3}\ \ &  \ R > R_{eq}
\end{array} \right \}
$$
The mass inside the Hubble radius, $M_H$ is similarly defined to be [3,6]
$$
M_H=\frac{4\pi}{3}\rho(\frac {d_H}{2})^3\propto
\left\{ \begin{array}{ll}
 R^{-1}\ \ \  & \ R < R_{eq}\\
\rm const. \ \ \ & \ R > R_{eq}
\end{array} \right\}
$$
In our present model $M_H$  evolves
as $t^{-1}$ in the early universe and remains constant during the matter dominated
epoch.
While $M_H\propto t$ in the
standard scenario, our prediction is that $M_H$ is constant in the matter
dominated phase. We see that in the radiation dominated phase $M_J\propto
M_H\propto t^{-1}$, so if initially  $M_J >M_H$ then no matter will be
formed in the radiation dominated phase but if $M_J<M_H$  matter is
likely to form in this phase. $M_J$ decreases as $R^{-3}$ (faster than the
standard scenario) and
$M_H= \rm const.$ in the matter dominated phase. This is unlike the
standard scenario where $M_J\propto M_H\propto t$ in the RD phase and
$M_J\propto M_H^{-1}\propto t^{-1}$ in the MD phase.
We remark that the present model is consistent with the CMBR since we have
in the RD phase $T\propto R^{-1}$.
In the standard scenario the Jeans mass at equilibrium between radiation
and matter is calculated to be [10]
\begin{equation}
M^s_J=\frac{4\pi}{3}(\frac{T_r}{G m_H})^{3/2}\frac{1}{\rho_{eq.}^{1/2}}\sim
10^5 \rm M_{\odot}
\end{equation}
Where $M^s_J$ is the standard model value, $T_r$ the temperature of radiation at
equilibrium, $\rho_{eq}$ the density of matter at equilibrium, and $m_H$ is the mass
of hydrogen atom.
In our scenario the Jeans mass
is \begin{equation}
M_J=(\frac{t_{eq}}{t_0})^{3/2}M^s_J\sim 10^{11}\rm M_{\odot}
\end{equation}
where we have used $t_{eq}=10^6$ year.
This is  a typical mass of a galaxy. The mass in the horizon
radius, $M_H$ at the equilibrium is given by, with eq.(10),
\begin{equation}
M_H=\frac{c^3t_{eq}}{G_{eq}}=\frac{c^3t_0}{G_0}\sim 10^{11}\rm M_{\odot}
\end{equation}
Thus we see that $M_J\sim M_H$ at equilibrium. We also notice that $M_H$ is
independent of time in the matter
dominated phase. Thus this mass is frozen on the onset of the
recombination time which is unlike the Jeans mass which still
decreases with time in both eras. This may show that
why we don't observe galaxies with  masses much bigger than this.
In fact, this mass is of the same order of the mass considered from
damping of photon viscosity, which was found to vary as $T^{-9/2}$
[10].
Thus a more quantitative analysis should be done before
one decides to take this model as an alternative to the standard scenario.
In fact our model behaves like a universe filled with string-like
matter with an equation of state  $p=-\frac{1}{3}\rho$ and with $G$
constant!
\\ \\
{\bf References}
\\ \\
1- A.I.Arbab, {\it gr-qc/9909044}\\
2- A.M.Abdel Rahman, {\it Gen. Rel. Gravit.22,655,1990}\\
3- T.Padmanabhan,{\it Structure formation in the Universe},CUP,1993\\
4- P.A.M.Dirac, {\it Proc. R.Soc. A165, 199(1838)\\
5- C.Brans and R.H.Dicke, {\it Phys. Rev.24},925, 1961\\
6- Ya.B.Zel'dovich and I.D.Novikov,{\it The structure and evolution of
the universe},UCP,1983\\
7- A.Beesham, {\it Phys. Rev48< 3539, 1993}\\
8- O.Bertolami, {\it Nouvo Cim. 93B, 36, 1986}\\
9- S.Ram and C.P.Singh, {\it Nouvo. Cim. 114B,45,1999}\\
10- M.Berry, {\it Principle of cosmology and gravitation}, Cambridge
university press (CUP), 1976\\
11- T.W.Sciama, {\it MNRAS}, 113,34,1953}, \ \ J.V.Narlikar, {\it Introduction to
cosmology}, CUP,1993, \\
12- A.I.Arbab, {\it gr-qc/9906045}\\
13- A.I.Arbab, {\it physics/9905049}\\
\newpage
\begin{center}
\begin{table}
\caption{Cosmological Parameters of the very Early Universe}
\vspace{1cm}
\begin{tabular}{|r|r|r|r|r|r|}
\hline
Parameter  & Planck(P)  & GUT(G) & Electroweak(W) &  Nuclear(N) & Present(0)\\
\hline
Time & $10^{-43}s$ & $10^{-40}s$ & $10^{-27}s$ & $10^{-23}s$ &
$10^{17}s$\\
\hline
Radius of the universe & $\rm 10^{-33}cm$ & $\rm 10^{-30}cm$ & $\rm
10^{-17}cm$ &
$\rm 10^{-13}cm$ & $\rm 10^{28}cm$\\
\hline
Strength of gravity & $G_0$ & $10^6 G_0$ & $10^{32}G_0$ & $10^{40}G_0$
& $G_0$ \\
\hline
Density of the universe & $\rm 10^{94}g cm^{-3}$ & $\rm 10^{62}g cm^{-3}$ &
$\rm 10^{26}g cm^{-3}$ & $\rm 10^{14}g cm^{-3}$ & $\rm 10^{-29} g cm^{-3}$\\
\hline
Mass Scale & $10^{19}\rm GeV$ & $10^{16}\rm GeV$ & $10^3\rm
GeV$ & $10^{-1}\rm GeV$ & $10^{80}\rm GeV$\\
\hline
\end{tabular}
\end{table}
\end{center}
\end{document}